\documentclass[preprint, amsmath,amssymb,aps,natbib,prf]{revtex4-2}
\usepackage{geometry}
\usepackage{comment}
\usepackage{enumerate}
\usepackage{amsmath,nccmath}
\usepackage{xcolor}
\usepackage{mathrsfs}
\usepackage{graphicx}
\usepackage{dcolumn}
\usepackage{yhmath}
\usepackage{natbib}
\usepackage{bm}

\usepackage[mathlines]{lineno}
\usepackage{xcolor}
\usepackage{color}
\usepackage[colorlinks=true,linktoc=page,citecolor=blue,linkcolor=blue,]{hyperref}

\newcommand{\bb}{{\bm b}}
\newcommand{\bu}{{\bm u}}

\newcommand{\bomega}{{\bm \omega}}
\newcommand{\ba}{{\bm a}}
\newcommand{\bQ}{{\bm Q}}

\newcommand{\bnabla}{{\bm \nabla}}

\begin{document}
\title{Exact scaling laws in isotropic binary fluid turbulence}

\author{Nandita Pan}
\email{nandita.pan08@gmail.com}
\affiliation{Department of Physics, Institute for Condensed Matter Physics,Technische Universität Darmstadt, Darmstadt 64289, Germany}
\affiliation{
 Department of Physics, Indian Institute of Technology Kanpur, Uttar Pradesh, 208016 India}

\author{Supratik Banerjee}
\email{sbanerjee@iitk.ac.in}
\affiliation{
 Department of Physics, Indian Institute of Technology Kanpur, Uttar Pradesh, 208016 India}

\date{\today}

\begin{abstract}
Binary fluid turbulence distinguishes itself from ordinary fluid turbulence by virtue of interfacial dynamics. Whether Kolmogorov-like scaling laws also exist for binary fluid turbulence is a fundamental question to explore. Starting from tensor formalism \`a la von K\'arm\'an and Howarth, here we derive exact scaling laws for isotropic Cahn-Hilliard-Navier-Stokes (CHNS) turbulence both in terms of two point correlators and increments. In particular, we derive the CHNS analogs for $1/3$, $4/3$, $2/15$ and $4/5$ laws known for isotropic hydrodynamic turbulence and show that the new scaling laws contain contributions both from the bulk flow and interface. The $2/15$ and $4/5$ laws of CHNS turbulence are found to be expressed purely in terms of two-point correlators and structure functions and their derivatives, respectively. However, unlike their hydrodynamic counterparts, these relations  involve additional contributions from non-longitudinal directions. By means of direct numerical simulations with up to $1024^3$ grid points, all the derived exact laws are numerically verified and the scale dependence of the cascade rates obtained from different exact laws are thoroughly compared. As one moves from the homogeneous (but not necessarily isotropic) divergence form to the isotropic $4/5$ form, the inertial range is found to shift towards larger scales with a comparatively flatter cascade rate profile as a result of successive integrations over the small scales.

\end{abstract}
\maketitle

\section{Introduction}
\label{sec:headings}
Despite the highly nonlinear complex character, turbulence is ubiquitous in nature and therefore demands a thorough understanding. A fully developed turbulent flow exhibits multi-scale behaviour associated with the scale-by-scale transfer of kinetic energy, an inviscid invariant of the flow, from larger to smaller scales. Across the intermediate inertial scales, where the large-scale energy injection and small-scale dissipation play little role, the kinetic energy cascades at a constant rate ($\varepsilon$), thus beckoning a universal feature of hydrodynamic turbulence. The knowledge of $\varepsilon$ is physically important as for stationary turbulence, it also equals to the energy injection and dissipation rates and therefore the turbulent heating rate. To calculate $\varepsilon$ in real space, one has to derive the evolution equation of the two-point correlators of velocity. 
For homogeneous and isotropic turbulence, such a framework was developed by 
\cite{von_Karman_1938}, where specific isotropic forms for the two-point correlation tensors are assumed \cite{Djenidi_Antonia_2022_VKH}. For a statistically stationary regime, the cascade rate of incompressible hydrodynamic (HD) turbulence can be exactly computed by Kolmogorov's famous $4/5$ law as $ S_3 = \langle \delta u^3_{r} \rangle = - 4/5 \varepsilon r$, where $\delta u_{r} = \delta \bm{u}\cdot \bm{\hat{r}}$ is the longitudinal component of $\delta \bu = \bu (\bm{x}+\bm{r}) - \bu (\bm{x})$, representing the two-point increment of the velocity field $\bu$ associated with the length scale $r$ and $\langle \cdot \rangle$ represents the statistical average \citep{kolmogorov1941c}. 
Equivalently, $\varepsilon$ can also be calculated in terms of two-point velocity correlators through the $2/15$ exact law given by $ \langle u^2_r(\bm{x})u_r(\bm{x+r}) \rangle = -2/15 \varepsilon r$ \cite{Batchelor_1953_Homogeneous_Turbulence}.

In contrast to pure hydrodynamics, separate exact laws for two-point increments and correlators cannot be derived for incompressible magnetohydrodynamics (MHD), and the MHD analog of the 4/5 law contains both two-point correlators and increments as $\langle \delta u^3_{r}\rangle - 6 \langle b^2_{r}(\bm{x}) u_{r}(\bm{x} + \bm{r})\rangle = -4/5 \varepsilon r $, where $\bm{b}(\bm{x})$ is the magnetic field and $\varepsilon$ is the total energy (kinetic + magnetic) transfer rate \citep{Politano_1998_vKH_MHD, Yoshimatsu_2012}. Similar exact laws describing the transfer of pseudo-energies $\langle |z^{\pm}|^2/2\rangle$ take the form $\langle \delta {z_r^{\pm}}^2 \delta z_r^{\mp} \rangle -2\langle {z_r^{\pm}}^2(\bm{x}) z^{\mp}_r(\bm{x} + \bm{r})) \rangle = - 8/15 \varepsilon^{\pm} r$, respectively where $\bm{z}^{\pm} = \bm{u} \pm \bm{b}$ denote the Els\"{a}sser variables. Beyond HD and MHD, von-Karman-Haworth (vKH) type exact relations are  derived for energy cascade in other turbulent flows e.g., HD in the presence of rotation, Hall-MHD, two-fluid plasmas \textit{etc.} \citep{Saga_Bhattacharjee_2007, Galtier2008HMHD_VonKarman, Andres_2016_VonKarman_twofluid}, and also for various helical invariants   \citep{Gomez_2000_Exact_helical_flows, Politano_2003_helical_MHD}.

Upon relaxing the assumption of isotropy, one can follow the Yaglom-Monin (YM) formulation to derive a divergence form  $\bnabla_{\bm{r}}\cdot\bm{\mathcal{F}} = -4 \varepsilon$, where the flux term $\bm{\mathcal{F}} = \langle (\delta \bu)^2 \delta \bu\rangle$ for HD turbulence and $\bm{\mathcal{F}} = \langle [(\delta \bu)^2 + (\delta \bb)^2] \delta \bu - 2 (\delta \bu \cdot \delta \bb) \delta \bb \rangle$ for MHD turbulence \citep{monin_Yaglom_1975}. If the assumption of isotropy is now included, we have $\bnabla_{\bm{r}} \cdot \mathcal{F} \equiv r^{-2} \partial /\partial_r(r^2 \mathcal{F}_r )  = -4 \varepsilon $. Integrating over a sphere of radius r, one finally obtains the popular $4/3$ forms of the exact laws as $\mathcal{F}_r = -4/3 \varepsilon r$ 
where $\mathcal{F}_r  = \langle (\delta \bu)^2 \delta u_r \rangle$ for HD turbulence, and    $ \mathcal{F}_r  = \langle [(\delta \bu)^2  + (\delta \bb)^2] \delta u_r - 2 (\delta \bu \cdot \delta \bb)) \delta b_r\rangle$
for MHD turbulence \citep{Politano_1998_dynamical_MHD}. Similar $4/3$ laws are derived for several other turbulent flows, including Hall and electron MHD  \citep{ Celani_1998, Ferrand_2019_HMD_incmop} and for cascades associated with helical invariants \citep{Politano_1998_dynamical_MHD, Podesta_2008_incompressible_magnetohydrodynamic_turbulence}. Interestingly, starting with vKH formalism, one can also obtain the $4/3$ laws \citep{antonia1997analogy}. However, using YM formulation, one cannot access the detailed partitioning between longitudinal and non-longitudinal contributions in the correlation tensors and hence the derivation of a tensorial $4/5$ law is not possible. This is also consistent with the fact that the existence of a $4/3$ law does not necessarily imply the existence of a hydrodynamic-like $4/5$ law where the cascade rate can be known exclusively in terms of the longitudinal increments (\textit{e.g.} the case of MHD turbulence). Note that, the case of isotropic passive scalar (a scalar field which does not affect the flow dynamics i.e. the temperature in an incompressible flow) turbulence is particular since no $4/5$ exact law exists for such a flow and one obtains a $4/3$ exact law given by $\langle  \delta \phi^2 \delta u_{r}\rangle = -4/3 \varepsilon_{\phi}r$, where $\phi$ is the scalar-field and $\varepsilon_{\phi}$ is the flux rate corresponding to the transfer of scalar energy $\langle\phi^2/2\rangle$. 

In addition to neutral fluids and plasmas, turbulence in binary fluids has attracted significant attention in recent years. Below a critical temperature ($T_c$), binary fluids undergo spinodal phase separation where fluid domains coarsen to minimize the interface between the two components \citep{Bray_1995_Theory_phase-ordering_kinetics, Cates_2018_phase-separation_kinetics_active_emulsions}. 
The presence of turbulence, however, prevents this process through nonlinear fragmentation of the fluid domains, subsequently leading to a phase-arrested state of emulsion which is crucial not only for the fundamental understanding of multiphase fluid dynamics but also for a wide range of commercial applications including froth formation, food preservation, cosmetic industries, \textit{etc.}  \citep{Hinze_1955, Berti_2005_Active_Passive_Binary_Mixtures, Perlekar_2014_spinodal_Isotropic_Turbulence, Wang_eFood_2023, Shrinivas_2008, Colucci_2020_Water-in-Oil_Emulsions_Delivery_Vehicles, Green_Scilight_crude_oil, Muzzio_1991_self-similar}. 

The principal difference between ordinary and binary fluid turbulence (BFT) arises from the presence of interfacial dynamics which actively couples back to the flow. Depending on how the interfaces are treated, binary fluids are typically modeled in two ways: (i) sharp-interface models, where the two fluids are separated by sharp (zero-thickness) interfaces (e.g., the volume of fluid method, front tracking method, level set methods, etc.) consisting of Navier-Stokes (NS) equation for the bulk flow with jump or boundary conditions imposed at the interfaces to ensure conservation of mass, momentum, and energy \citep{Rajkotwala_20191_sharpInterfaceModels, Thiesset_Vahe_2025, Esposito_2023_droplet_dynamics_turbulence_cascade}, and (ii) diffuse-interface models \citep{SEKERKA_2004}, 
where the interface between the two immiscible fluids is described by a continuously varying transition layer having finite width and defined through the variation of a composition or phase-field \( \phi \) \citep{Bray_1995_Theory_phase-ordering_kinetics, Jacqmin_1999_Two-PhaseNavierStokesFlowsPhase-FieldModeling, Cates_2018_phase-separation_kinetics_active_emulsions}. 

 Within the diffused interface model, BFT is often studied in the framework of Cahn-Hilliard-Navier-Stokes (CHNS) equations, in which the interfacial feedback is modeled by means of Korteweg stress given by $\xi \boldsymbol{\nabla} \cdot \left({\bf Q} \otimes {\bf Q} \right)$, where $\xi$ is the interfacial width parameter, $\bQ = \bnabla  \phi$ is the composition gradient field, $\phi = (\rho_A - \rho_B)/(\rho_A + \rho_B)$ and $\rho_A$ and $\rho_B$ represents the densities of the two-fluids, respectively \citep{Bray_1995_Theory_phase-ordering_kinetics, Berti_2005_Active_Passive_Binary_Mixtures, Perlekar_2019_Kinetic_BFT,  Roccon_phase_field_2023}. Unlike Navier-Stokes turbulence in ordinary fluids, the kinetic energy is no longer conserved in CHNS flow, and a clear departure from an HD-like kinetic energy cascade is observed in phase arrested turbulent binary fluids \citep{Perlekar_2014_spinodal_Isotropic_Turbulence, Perlekar_2019_Kinetic_BFT}.  
Such flows, on the other hand, admit a different inviscid invariant given by the 
sum of the kinetic and the interfacial energies expressed as $\langle u^2/2 +\xi Q^2/2 \rangle$ \citep{Ruiz_1981_BFT, Bhattacharjee_2021_driven_active_matter}. Recently, for homogeneous CHNS turbulence, we derived a YM-type exact law given by \citep{Pan_2022_Exact_BFT} 
\begin{equation}
 \left\langle \left[(\delta \bu)^2 - \xi (\delta \bQ)^2 \right]\delta \bu
+ 2 \xi (\delta \bu \cdot \delta \bQ)\delta \bQ \right\rangle  = - 4 \varepsilon. \label{diveregnce_form}
\end{equation} Using high-resolution direct numerical simulations (DNSs) (with up to $1024^3$ grid points), we clearly show the existence of an inertial range cascade for the total energy characterized by a constant $\varepsilon$.  A natural question that then arises is whether a CHNS turbulence sustains a Kolmogorov-like $4/5$ law. Despite the aforementioned systematic studies, no dedicated study has addressed this issue till date. At first glance, the structural similarity between the YM exact laws for MHD and CHNS turbulence might look helpful to conclude over the possibility of a $4/5$ law in BFT.  However, unlike the magnetic field, the composition gradient field, $\bQ = \bnabla \phi$ is a curl-free vector and hence the form of two-point correlation tensors involving $\bQ$  will be different and a separate dedicated study is therefore needed to explore the possibility of a vKH-like $4/5$ law in CHNS turbulence. 

In this paper, using vKH type isotropic tensor formalism, we rigorously derive the BFT analogue of the isotropic exact laws for the total energy cascade both in terms of two-point correlators and increments. Unlike incompressible HD and MHD turbulence, the combination of solenoidal ${\bu}$ and irrotational ${\bQ}$ renders the analysis considerably challenging and non-trivial. Using DNS with $512^3$ and $1024^3$ grid points, we numerically investigate the universality of the derived exact laws. Furthermore, we compare both the $4/3$ and $4/5$ laws with their counterparts in the correlator form. Finally, we discuss the variation of the width of the inertial range according to the chosen form of the exact laws.

The paper is organized as follows. In Sec.~\ref{derivation_of_exact_laws}, using the tensor formalism, we derive the evolution equations of the two-point correlators in terms of the isotropic tensors, and subsequently obtain the various forms of the isotropic exact laws both in terms of two-point correlators and two-point increments. In Sec.~\ref{numerical_investigation_of_exact_laws}, we describe the numerical methods, simulation details and flux computation techniques followed by a systematic discussion of the obtained results along with a comparison between the derived exact laws. Finally, we conclude in Sec.~\ref{summary_and_conclusion} with a summary and possible future directions.

\section{Derivation of the vKH equations for CHNS turbulence}\label{derivation_of_exact_laws}

In the incompressible CHNS model, the complete set of evolution equations for the dynamical variables $\bu$ and $\bQ$ is given by \citep{Pan_2022_Exact_BFT, Pan_2024_Relaxation_BFT} 
 \begin{gather}
     \partial_{t} \bu =  -(\bm{u}\cdot\bnabla) \bm{u} -   \xi \bQ (\bnabla \cdot \bQ) - \bnabla P + \nu \nabla^2 \bu + \bm{f} \label{Eu},\\
     \partial_{t} \bm{Q} =  -\bnabla(\bu\cdot \bQ) + \mathcal{M} \nabla^2 \bnabla\mu , \label{EQ}
 \end{gather}
with $\bnabla\cdot\bu =0$ and $\bnabla \times\bQ=\bm{0}$, $\bm{f}$ is a large-scale turbulent forcing, $\nu$ is the kinematic viscosity, $\mathcal{M}$ is the mobility coefficient, $P= p  + \phi \mu - \phi^2/2 - \phi^4/4$ represents the total pressure with $p$ being the fluid pressure, $\mu = \delta \mathcal{F}/\delta \phi = -\phi + \phi^3 - \kappa \nabla^2 \phi$ the chemical potential and $\kappa = \xi \rho = \xi$ for 
 $\rho (= \rho_A + \rho_B) =1 $. The numerical values of the parameters are listed in Tab.~\ref{Table1}.
The vKH equation connects the time evolutions of the two-point second-order correlators of the inviscid invariants to the corresponding third-order correlators. The two-point correlator for the total energy is given by
\begin{equation}
    \mathcal{R}_E = \left\langle \frac{\bu \cdot \bu' + \xi \bQ\cdot\bQ'}{2}\right\rangle,
\end{equation}
where the primed and unprimed quantities represent the flow fields at space points $\bm{x}$ and $\bm{x}' = \bm{x}+\bm{r}$, respectively. Using ~\eqref{Eu}--\eqref{EQ}, the time evolution of $\mathcal{R}_E$ can be written as \citep{Pan_2022_Exact_BFT}
\begin{align}
 \partial_t \mathcal{R}_E  = & \frac{1}{2}\bnabla_{\bm{r}}\cdot \left\langle \bu ( \bm{u}\cdot \bm{u}^\prime) - \bu' ( \bm{u}\cdot \bm{u}^\prime) + \xi[(\bm{u}^\prime\cdot \bm{Q})\bm{Q}-( \bm{u}\cdot\bm{Q}^\prime)\bm{Q}^\prime \right.\nonumber\\
& +\left.(\bm{u}\cdot\bm{Q})\bm{Q}^\prime - (\bm{u}^\prime\cdot\bm{Q}^\prime) \bm{Q}]\right\rangle + \mathcal{D}(\bm{r}) + \mathcal{F}(\bm{r}), \label{RE_F1}
 \end{align}
where $\mathcal{D}(\bm{r}) = \mathcal{D}_u (\bm{r}) + \mathcal{D}_Q(\bm{r}) $ is the average two-point total energy dissipation rate with $\mathcal{D}_u(\bm{r}) = \nu \nabla^2_{\bm{r}}\langle  \bu\cdot\bu' + \bu'\cdot\bu\rangle/2 $ and $\mathcal{D}_Q (\bm{r}) = \xi \mathcal{M}\nabla^2_{\bm{r}}\langle \bQ \cdot \bnabla' \mu' + \bQ' \cdot \bnabla \mu \rangle /2 $ representing the average two-point kinetic and active energy dissipation rates, respectively and $\mathcal{F}(\bm{r})= \langle \bm{f}\cdot \bm{u}' + \bm{f}'\cdot\bm{u} \rangle/2$ is the average two-point total energy injection rate. Assuming isotropy and homogeneity, we define the two-point second-order isotropic tensors for $\bu$ and $\bQ$ fields as \citep{Batchelor_1953_Homogeneous_Turbulence}
\begin{align}
   R^{1}_{ij} (\bm{r}) &= \langle u_i u'_j \rangle = F_1 r_i r_j + G_1 \delta_{ij}, \label{R_uu}\\
 R^{2}_{ij} (\bm{r}) &= \langle Q_i Q'_j \rangle = F_2 r_i r_j + G_2 \delta_{ij}, \label{R_QQ}
\end{align}
where $F_{1,2}\equiv F_{1,2}(r)$ and $G_{1,2}\equiv G_{1,2}(r)$ are arbitrary functions of $r$. Using the divergence-free property of $\bu$ and statistical homogeneity, we get  $ \langle\bnabla\cdot (\bu \otimes\bu')\rangle =\bm{0} \implies \partial_{r_j} R^{1}_{ij} (\bm{r}) = \bm{0}$, leading to a relation between $F_{1}(r)$ and $G_{1}(r)$  as
\begin{align}
    \hspace{-0.3cm}\left(r\frac{\partial F_1}{\partial r} + 4 F_1 + \frac{1}{r} \frac{\partial G_1}{\partial r}\right) r_i = 0 \,\,\,\,\ \implies \,\,\,\, \frac{\partial}{\partial r} (F_1 r^2 + G_1) = -2 F_1 r. \label{Ruu_divergence_free}
\end{align}
Here we used partial derivatives $ \partial F_{1}(r)/\partial r$ and $ \partial G_{1}(r)/\partial r$ instead of total derivatives, since $F_1$ and $G_1$ are functions of time as well. Unlike the velocity field $\bu$, the composition gradient $\bQ$ is an irrotational field, which introduces an important difference in the analysis of tensors involving $\bQ$.  In this case, we note $ \langle \bQ \otimes (\bnabla' \times\bQ' ) \rangle = \bm{0}$ or equivalently $\langle Q_i \epsilon_{\alpha \beta j}\partial'_{\beta}Q'_j \rangle = \langle \epsilon_{\alpha \beta j}\partial'_{\beta}( Q_iQ'_j) \rangle=  \epsilon_{\alpha \beta j}\partial_{r_{\beta}} R^{2}_{ij} (\bm{r}) = 0$, which further gives
\begin{equation}
        \epsilon_{\alpha \beta i} r_{\beta} \left(-F_2 +  \frac{1}{r}\frac{\partial G_1}{\partial r}\right) = 0  \,\,\,\, \implies \,\,\,\, F_2 =  \frac{1}{r}\frac{\partial G_1}{\partial r}. \label{RQQ_curl_free}
\end{equation}

\noindent Next, we define the longitudinal and lateral two-point correlators as 
\begin{align}
  \hspace{-0.3cm}  R^{1}_{rr} = \langle u_{r} u'_{r} \rangle = F_1 r^2 + G_1 \,\,\text{and} \,\, R^{1}_{n n} = \langle u_{n} u'_{n} \rangle = G_1, \label{Ruu_LL}\\
  \hspace{-0.2cm} \hspace{-0.3cm}  R^{2}_{r r} = \langle Q_{r} Q'_r \rangle = F_2 r^2 + G_2 \,\,\text{and} \,\, R^{2}_{n n} = \langle Q_{n} Q'_{n} \rangle = G_2,\label{RQQ_LL}
\end{align}
\noindent where for a vector field ${\boldsymbol a}$,  $a_r = \bm{a}\cdot\bm{\hat{r}}$ and $a_n = \ba\cdot\bm{\widehat{n}}$ ( $\bm{\widehat{n}}$ representing any one direction perpendicular to $\bm{\widehat{r}}$) represent the longitudinal and lateral components, respectively, and by definition, $r_r \equiv r,\, r_n = 0$ and $\delta_{rr} = \delta_{nn} =1 $\citep{Batchelor_1953_Homogeneous_Turbulence,landau_1987fluid}. Substituting ~\eqref{Ruu_LL} and \eqref{RQQ_LL} in \eqref{R_uu} and \eqref{R_QQ}, we obtain 
\begin{align}
    R^{1}_{ij}(\bm{r}) &= \frac{r}{2}\frac{\partial }{\partial  r} \left(R^{1}_{rr}\right) \left( \delta_{ij} - \frac{r_i r_j}{r^2} \right) + R^{1}_{rr} \delta_{ij},\label{R_uu_LL}\\
    R^{2}_{ij}(\bm{r}) &= \frac{1}{r}\frac{\partial }{\partial  r}\left( R^{2}_{nn}\right)  r_i r_j + R^{2}_{nn} \delta_{ij} \label{R_QQ_NN}.
\end{align}
Therefore, we have $\langle \bu \cdot \bu' \rangle =  R^{1}_{ii} = r^{-2} \partial \left(r^3 R^{1}_{r r}\right)/\partial r$ and $\langle \bQ \cdot \bQ' \rangle =  R^{2}_{ii} = r^{-2} \partial \left(r^3 R^{2}_{nn}\right)/\partial r$ and hence the l.h.s. of equation ~\eqref{RE_F1} becomes
\begin{equation}
    \partial_t \mathcal{R}_E = \frac{1}{2r^2}\frac{\partial }{\partial  r}\left[r^3 \left( \partial_t R^1_{rr} + \xi \partial_t R^2_{nn}\right) \right].\label{del_T_R_LL_NN}
\end{equation}

\noindent Proceeding further, we define the following third-order isotropic tensors \citep{Batchelor_1953_Homogeneous_Turbulence, Robertson_1940} 

\begin{align}
     \mathcal{C}^{1}_{ijm}(\bm{r}) &= \langle u_i u_j u'_m\rangle = A_1r_i r_j r_m + B_1 (r_i \delta_{jm} + r_j \delta_{im}) + D_1 r_m \delta_{ij}, \label{C1_ijm}\\
     \mathcal{C}^{2}_{ijm}(\bm{r}) &= \langle Q_i Q_j u'_m\rangle 
    = A_2r_i r_j r_m + B_2 (r_i \delta_{jm} + r_j \delta_{im}) + D_2 r_m \delta_{ij}, \label{C2_ijm}\\
    \mathcal{C}^{3}_{ijm}(\bm{r}) &= \langle u_i Q_j Q'_m\rangle = A_3r_i r_j r_m + B_3 r_i \delta_{jm} + C_3 r_j \delta_{im} + D_3 r_m \delta_{ij}\label{C3_ijm}, 
\end{align}
where $A_{1,2,3} \equiv A_{1,2,3}(r)$, $B_{1,2,3} \equiv B_{1,2,3}(r)$, $C_3 \equiv C_3(r)$ and $D_{1,2,3} \equiv D_{1,2,3}(r)$ are the arbitrary functions of $r$. 
Unlike $\mathcal{C}^{1,2}_{ijm}$, $\mathcal{C}^3_{ijm}$ is not symmetric in the indices $i$ and $j$, and hence a different coefficient $C_3$ is introduced. 
Using the expressions, one can write
\begin{align}
    \langle  u_r (\bu\cdot\bu')\rangle &= \mathcal{C}^1_{rjj} = \mathcal{C}^1_{rrr} + 2 \mathcal{C}^1_{rnn}, \label{C1rjj}\\
    \langle Q_r (\bQ \cdot\bu')\rangle &= \mathcal{C}^2_{rjj} = \mathcal{C}^2_{rrr} + 2 \mathcal{C}^2_{rnn},\label{C2rjj}\\
    \langle (\bu\cdot\bQ) Q_r' \rangle & = \mathcal{C}^3_{iir}\, = \mathcal{C}^3_{rrr} + 2 \mathcal{C}^3_{nnr}.\label{C3iir}
\end{align}
Under the assumption of isotropy, the $\theta$ and $\phi$ components of the divergence operator in the r.h.s. of the equation \eqref{RE_F1} can be discarded, which gives

\begin{align}
    \partial_t \mathcal{R}_E &= \frac{1}{r^2}\frac{\partial }{\partial  r}\left( r^2 \left\langle  u_r (\bu\cdot\bu') + \xi Q_r (\bu'\cdot\bQ) +  \xi Q_r'(\bu\cdot\bQ)\right\rangle \right) + \mathcal{D}(r)+ \mathcal{F}(r),\nonumber \\
    &=\frac{1}{r^2}\frac{\partial }{\partial  r}\left( r^2 \left[\mathcal{C}^{1}_{rrr} +\xi( \mathcal{C}^{2}_{rrr}+ \mathcal{C}^{3}_{rrr})+ 2 \mathcal{C}^{1}_{rnn} + 2\xi (\mathcal{C}^2_{rnn}+\mathcal{C}^3_{nnr})\right] \right) + \mathcal{D}(r)+ \mathcal{F}(r),\label{del_RT_r_plus_diss}
\end{align}
where $\mathcal{D}(r)$ and $\mathcal{F}(r)$ are the isotropic contributions of $\mathcal{D}(\bm{r})$ and $\mathcal{F}(\bm{r})$, respectively. In the above, we used the anti-symmetry property (under $\bm{r}\rightarrow -\bm{r}$) of the third-order isotropic tensors defined in equations ~\eqref{C1_ijm}-~\eqref{C3_ijm}. Finally, equating the expressions of $\partial_t \mathcal{R}_E$ from equations \eqref{del_T_R_LL_NN} and \eqref{del_RT_r_plus_diss}, and integrating them, we get 
\begin{align}
    \partial_t \left(R^1_{rr} + \xi R^2_{nn}\right) =  \frac{2}{r} \mathcal{A} +\frac{2}{3}\left(D_{iso_{1}}+F_{iso_{1}} \right), \label{vKH_BFT}
\end{align}
where $\mathcal{A} = \left[\mathcal{C}^{1}_{rrr} +\xi (\mathcal{C}^{2}_{rrr}+ \mathcal{C}^{3}_{rrr})+ 2 \mathcal{C}^{1}_{rnn} + 2\xi (\mathcal{C}^2_{rnn}+\mathcal{C}^3_{nnr})\right]$, $\mathcal{D}_{iso_1} = 3 r^{-3}\int \mathcal{D}(r)r^2 dr$ and $\mathcal{F}_{iso_1} = 3 r^{-3}\int \mathcal{F}(r)r^2 dr$. In the above, we have set the constant of integration to zero to avoid any singular solution at $r=0$. Eq.~\eqref{vKH_BFT} is the vKH equation corresponding to total energy transfer in BFT. 

\subsection{Derivations of $1/3$ and $4/3$ exact laws}

(i) \textbf{\textit{In terms of two-point correlators}}: Under the assumption of statistical stationarity, l.h.s of the Eq.~\eqref{vKH_BFT} vanishes. In the limit of infinite Reynolds number, the dissipative term $D_{iso_1}$ can be neglected inside the inertial range. $\bm{f}$ being the large-scale forcing, the two-point average injection rate inside the inertial range becomes $\mathcal{F}(r)=\langle \bm{f}\cdot\bm{u}'+\bm{f}'\cdot\bm{u}\rangle/2 \approx \langle \bm{f}\cdot \bm{u} \rangle = \mathcal{F}(0) \equiv\varepsilon$, where $\varepsilon$ is the constant energy injection rate, leading to $\mathcal{F}_{iso_1}(r) = \frac{3}{r^3}\int \mathcal{F}(r) r^2 dr = \varepsilon$. Combining all these, the final exact relation becomes
\begin{align}
\mathcal{C}^{1}_{rrr} + \xi \left(\mathcal{C}^{2}_{rrr}+ \mathcal{C}^{3}_{rrr}\right) + 2 \left[\mathcal{C}^{1}_{rnn} + \xi\left(\mathcal{C}^{2}_{rnn}+ \mathcal{C}^3_{nnr}\right)\right] &=  - \frac{1}{3}\varepsilon r\label{1_3law}.
\end{align}
Eq.~\eqref{1_3law} is the first main result of this paper. It represents an exact law for isotropic CHNS turbulence expressed in terms of mixed longitudinal and lateral two-point correlators. \\

\noindent (ii) \textbf{\textit{In terms of structure functions}}:
To cast the above exact relation in terms of two-point fluctuations or the structure functions of the field variables, we define the following isotropic structure functions
\begin{align}
\mathcal{B}^{1}_{ijm} &= \langle\delta u_i \delta u_j \delta u_m\rangle = 2(\mathcal{C}^1_{ijm} + \mathcal{C}^1_{imj} + \mathcal{C}^1_{jmi}) \label{B_1_ijm},\\
    \mathcal{B}^{2}_{ijm} &= \langle\delta u_i \delta Q_j \delta Q_m\rangle = 2(\mathcal{C}^3_{ijm} + \mathcal{C}^3_{imj} + \mathcal{C}^2_{jmi}) \label{B_2_ijm},
\end{align}
where $\delta \bm{a} = \bm{a}'-\bm{a}$ for any field variable $\bm{a}$. In obtaining above, the antisymmetry property  of the tensors $C^{1,2,3}_{ijm}(\bm{r})$ are used i.e., $C^{1,2,3}_{ijm}(\bm{r}) = - C^{1,2,3}_{ijm}(-\bm{r})$. Furthermore, using statistical homogeneity and incompressibility, it is straightforward to show $\langle \bm{\nabla}' \cdot [\bu' \otimes(\bu\otimes\bu)]\rangle = \bm{0}$ and $\langle \bnabla' \cdot [ \bu' \otimes (\bQ\otimes\bQ)]\rangle = \bm{0}$, leading to $\partial_{r_m} \mathcal{C}^{1,2}_{ijm} (\bm{r}) = \bm{0}$, which gives 
\begin{align}
    \frac{\partial A_{1,2}}{ \partial r} \frac{r_i r_j r^2_m}{r} +  A_{1,2} (r_j r_m \delta_{im} +  r_i r_m \delta_{jm} + & r_i r_j \delta_{mm}) +  \frac{\partial B_{1,2}}{ \partial r}\left(\frac{r_i r_m}{r} \delta_{jm} + \frac{r_j r_m}{r} \delta_{im}\right) \nonumber\\
    + 2B_{1,2}\delta_{im}\delta_{jm} &+ \frac{\partial D_{1,2}}{ \partial r}\frac{r^2_m}{r} \delta_{ij} + D_{1,2} \delta_{mm}\delta_{ij} = 0,
\end{align}
or equivalently 
\begin{equation}
      \left(r \frac{\partial A_{1,2}}{ \partial r}  + 5 A_{1,2} + 2 \frac{1}{r}\frac{\partial B_{1,2}}{ \partial r}\right) r_i r_j + \left( 2B_{1,2} + 3 D_{1,2} + r \frac{\partial D_{1,2}}{ \partial r}\right) \delta_{ij}= 0.
\end{equation}
For the above to be true for every $i $ and $j$, we get 

\begin{align}
r^2 \frac{\partial A_{1,2}}{\partial r} + 5 r A_{1,2} + 2 \frac{\partial B_{1,2}}{\partial r} &= 0,\,\,\text{and}\label{i_equal_j}\\
    2B_{1,2} + 3 D_{1,2} + r \frac{\partial D_{1,2}}{\partial r} &= 0.\label{i_not_equal_j}
\end{align}
Substituting $2B_{1,2}$ from Eq.~\eqref{i_not_equal_j} into Eq.~\eqref{i_equal_j}, we get
\begin{align}
    r^2 \, \frac{\partial A_{1,2}}{\partial r} + 5r A_{1,2} - \frac{\partial }{\partial r}\left(3D_{1,2} + r \, \frac{\partial D_{1,2}}{\partial r}\right) 
    &= 0 \nonumber \\
    \implies \quad
    \frac{1}{r^3} \left( r^5 \frac{\partial A_{1,2}}{\partial r} + 5r^4 A_{1,2} - 4r^3 \frac{\partial D_{1,2}}{\partial r} - r^4 \frac{\partial^2 D_{1,2}}{\partial r^2} \right) 
    &= 0 \nonumber \\
     \implies \quad 
    \frac{\partial }{\partial r}\left( r^5 A_{1,2}  -  r^4 \frac{\partial   D_{1,2}}{\partial r} \right) 
    &= 0
\end{align}
Integrating above w.r.t. $r$ and avoiding singular solution at $r=0$ (by noting that  $A_{1,2}$ and $D_{1,2}$ should remain finite at $r= 0$), we obtain
\begin{equation}
 A_{1,2} = \frac{1}{r}\frac{\partial D_{1,2}}{\partial r} \label{A_12_D_12}
    \end{equation} 
Again, we have $\langle (\bu \otimes \bQ)\otimes (\bnabla'\times\bQ')\rangle = \bm{0}\implies \langle u_i Q_j \epsilon_{\alpha \beta m} \partial'_{\beta} Q'_m\rangle= \epsilon_{\alpha \beta m}\partial_{r_{\beta}} \mathcal{C}^{3}_{ijm} (\bm{r}) = 0 $, thus leading to 
\begin{gather}
      \epsilon_{\alpha \beta m} \left[ \frac{\partial  A_{3}}{\partial r} \frac{r_i r_j r_m r_{\beta}}{r} +  A_{3} (r_j r_m \delta_{i \beta} +  r_i r_m \delta_{j \beta} +  r_i r_j \delta_{m \beta}) +  \frac{\partial  B_{3}}{\partial r}\frac{r_i r_\beta}{r} \delta_{jm} + \frac{\partial  C_{3}}{\partial r}\frac{r_j r_\beta}{r} \delta_{im} \right.\nonumber\\
    + \left. B_{3}\delta_{i \beta}\delta_{jm}  + C_{3}\delta_{j \beta}\delta_{im} + \frac{\partial  D_{3}}{\partial r}r_m r_\beta \delta_{ij} + D_{3} \delta_{m \beta}\delta_{ij} \right]= 0, \nonumber\\
    \implies \epsilon_{\alpha j m}r_i r_{m}\left(A_3- \frac{1}{r}\frac{\partial B_{3}}{\partial r} \right) + \epsilon_{\alpha i m}r_j r_{m}\left(A_3- \frac{1}{r}\frac{\partial C_{3}}{\partial r} \right) + \epsilon_{\alpha i j} (B_3 - C_3) = 0.\label{A3_curl_free2}
\end{gather}
Equation~\eqref{A3_curl_free2} has to be true for every combination of $\alpha, i$ and $j$. By taking $\alpha=i=1$ and $j=2$, one can show $\epsilon_{123}r_1r_3[A_3-(\partial B_3/\partial r)/r] = 0 \implies A_3 =(\partial B_3/\partial r)/r $. Similarly, for $\alpha=j=1$ and $i=2$, we get $\epsilon_{123}r_1r_3[A_3-(\partial C_3/\partial r)/r] = 0 \implies A_3 =(\partial C_3/\partial r)/r $. These two conditions also imply $B_3 = C_3$. Therefore, combining all, we get   
\begin{equation}
A_3= \frac{1}{r}\frac{\partial B_{3}}{\partial r} \,\,\text{and}\,\, C_{3} = B_{3}.\label{A3_B3_C3_D3}
\end{equation}
\noindent Using the relations~\eqref{A_12_D_12} and \eqref{A3_B3_C3_D3} in equations~\eqref{C1_ijm}-\eqref{C3_ijm}, we further obtain
\begin{align}
     \mathcal{C}^{1,2}_{ijm}(\bm{r}) &=
     \frac{1}{r}\frac{\partial  D_{1,2}}{\partial r} r_i r_j r_m + B_{1,2} (r_i \delta_{jm} + r_j \delta_{im}) + D_{1,2} r_m \delta_{ij}, \label{C1,2}\\
    \mathcal{C}^{3}_{ijm}(\bm{r}) 
    &= \frac{1}{r}\frac{\partial  B_{3}}{\partial r}r_i r_j r_m + B_3 (r_i \delta_{jm} + r_j \delta_{im}) + D_3 r_m \delta_{ij}\label{C3}. 
\end{align}

\noindent From the expressions above, it is straightforward to see
\begin{align}
    &\mathcal{C}^{1,2}_{r r r} = \frac{\partial  D_{1,2}}{\partial r} r^2 + 2B_{1,2} r + D_{1,2} r = -2 D_{1,2}r  \label{C_rrr12},\\
    &\mathcal{C}^{3}_{r r r} = \frac{\partial  B_{3}}{\partial r} r^2 + 2B_{3} r + D_{3} r, \label{C_rrr3}\\
   & \mathcal{C}^{1,2,3}_{ r n n} = \mathcal{C}^{1,2,3}_{ n  r n} =  B_{1,2,3}r, \label{C_rnn_123}\\
    &\mathcal{C}^{1,2,3}_{n n r} = D_{1,2,3} r, \label{C_nnr_123}
\end{align}
Combining Eqs.~\eqref{C_rrr12} and \eqref{C_nnr_123}, we get 
\begin{equation}
   \mathcal{C}^{1,2}_{r r r} + 2 \mathcal{C}^{1,2}_{n n r}   =0. \label{C12_rrr_C12_nnr}   
\end{equation}

\noindent Using the above equations and relations \eqref{C_rrr12}--\eqref{C12_rrr_C12_nnr}, one can write
\begin{align}
     \mathcal{B}^{1}_{rrr} &= \langle \delta u_r ^3\rangle = 6\mathcal{C}^1_{rrr} ,\label{B1_rrr}\\
     \mathcal{B}^{1}_{rnn} &= \langle\delta u_r\delta u^2_n\rangle = 2(2\mathcal{C}^1_{rnn} + \mathcal{C}^1_{nnr}),\label{B1_rnn}\\
     \mathcal{B}^{2}_{rrr} &= \langle\delta u_r\delta Q^2_r\rangle = 2(2\mathcal{C}^3_{rrr} + \mathcal{C}^2_{rrr}),\label{B2_rrr}\\
    \mathcal{B}^{2}_{rnn} &= \langle\delta u_r\delta Q^2_n\rangle = 2(2\mathcal{C}^3_{rnn} + \mathcal{C}^2_{nnr}),\label{B2_rnn}\\
    \mathcal{B}^{2}_{nnr} &= \langle\delta u_n \delta Q_n \delta Q_r \rangle =  2(\mathcal{C}^3_{nnr} + \mathcal{C}^3_{rnn} + \mathcal{C}^2_{rnn}) .\label{B2_nnr}
\end{align}
Rewriting the l.h.s of Eq.~\eqref{1_3law} in terms of the  structure functions, we finally obtain 
\begin{align}
(\mathcal{B}^1_{rrr}+\xi\mathcal{B}^2_{rrr}) + (2 \mathcal{B}^1_{rnn} - 2\xi\mathcal{B}^2_{rnn} + 4\xi \mathcal{B}^2_{nnr} ) &= -\frac{4}{3} \varepsilon r,
    \label{Exact_fluctuations}
\end{align}
which is the 4/3 exact law for the total energy transfer in CHNS turbulence. Unlike Eq.~\eqref{1_3law}, here a linear combination of longitudinal, lateral, and mixed structure functions is found to follow a linear scaling with the fluctuation length scale $r$. \\

\noindent Based on the Eqs.~\eqref{B_1_ijm}--\eqref{B_2_ijm} and \eqref{C_rrr12}--\eqref{B2_nnr}, one can show
\begin{align}
    \mathcal{B}^1_{rjj} &= \langle\delta u_r (\delta \bu)^2\rangle  =  2\left( 2\mathcal{C}^1_{rjj} + \mathcal{C}^1_{jjr} \right) = \mathcal{B}^1_{rrr} + 2\mathcal{B}^1_{rnn}\label{B1rjj},\\
     \mathcal{B}^2_{rjj} &= \langle\delta u_r (\delta \bQ)^2\rangle = 2\left( 2\mathcal{C}^3_{rjj} + \mathcal{C}^2_{jjr} \right) = \mathcal{B}^2_{rrr} + 2\mathcal{B}^2_{rnn} \label{B2rjj},\\
      \mathcal{B}^2_{jjr} &= \langle(\delta \bu \cdot \delta \bQ) \delta Q_r \rangle = 2\left( \mathcal{C}^3_{jjr} + \mathcal{C}^3_{jrj} + \mathcal{C}^2_{jrj}\right) = \mathcal{B}^2_{rrr} + 2\mathcal{B}^2_{nnr} \label{B2jjr},
\end{align}
and hence the Eq.~\eqref{Exact_fluctuations} can be expressed in a compact form as
\begin{align}
   \mathcal{B}^1_{rjj} - \xi \mathcal{B}^2_{rjj} + 2\xi \mathcal{B}^2_{jjr} &= - \frac{4}{3} \varepsilon r.\label{4_3rd_law}
\end{align}
This expression of the $4/3$ exact law can also be obtained by directly integrating the divergence form of the exact law ~\eqref{diveregnce_form} under the assumption of isotropy, where the divergence operator becomes spherically symmetric as, 
\begin{align}
    \frac{1}{r^2}\frac{\partial }{\partial r}\left[r^2\left(\mathcal{B}^1_{rjj} - \xi \mathcal{B}^2_{rjj} + 2\xi \mathcal{B}^2_{iir}\right) \right] = - 4 \varepsilon.\label{divergence_form_epsilon}
\end{align}
However, as we show below, obtaining Eq.~\eqref{Exact_fluctuations} (or \eqref{4_3rd_law}) from the divergence form is not helpful for deriving the $4/5$ law, for which, it is indeed essential to use the tensorial approach from the very beginning. 

\subsection{Obtaining $2/15$ and $4/5$ exact laws} 

In the next step, using the relations \eqref{C_rrr12}-\eqref{C_nnr_123}, we derive 
\begin{align}
         2 \mathcal{C}^{1,2}_{rnn} &=\mathcal{C}^{1,2}_{r rr} +\frac{r}{2}\frac{\partial \mathcal{C}^{1,2}_{rrr}}{\partial r}, \\
        \mathcal{C}^{3}_{nnr} &= \mathcal{C}^{3}_{r rr} -   \frac{\partial }{\partial r}(\mathcal{C}^{3}_{rnn}  r),
\end{align}
and with the help of these relations, we further write 
\begin{align}
   \mathcal{A}_1 &= \left(\mathcal{C}^{1}_{r r r} + 2 \mathcal{C}^{1}_{ r n n}\right) + \xi\left(\mathcal{C}^{2}_{r r r} + 2 \mathcal{C}^{2}_{ r n n}\right) + \xi\left(\mathcal{C}^{3}_{r r r} + 2 \mathcal{C}^{3}_{n n r}\right) \nonumber,\\
     &= 2\mathcal{C}^{1}_{r r r} + \frac{r}{2}\frac{ \partial}{\partial r}\left(\mathcal{C}^{1}_{r r r}\right) + \xi\left[2\mathcal{C}^{2}_{r r r} + \frac{r}{2}\frac{ \partial}{\partial r}\left(\mathcal{C}^{2}_{r r r}\right) + 4\mathcal{C}^{3}_{r r r} + r\frac{ \partial}{\partial r}\left(\mathcal{C}^{3}_{r r r}\right)\right]  \nonumber \\
     &+ \xi  \left[ 2\mathcal{C}^{3}_{n n r}-3\mathcal{C}^{3}_{r r r} - r\frac{ \partial}{\partial r}\left(\mathcal{C}^{3}_{r r r}\right)\right],\nonumber\\
    &= \frac{1}{2r^3}\frac{\partial }{\partial r}\left( r^4 \mathcal{C}^1_{r r r}\right)  + \frac{ \xi}{2r^3}\frac{\partial }{\partial r}\left[ r^4 \left(\mathcal{C}^2_{r r r} + 2 \mathcal{C}^3_{r r r}\right)\right] - \xi \frac{ \partial}{\partial r}\left[r\left( \mathcal{C}^3_{rrr} + 2\mathcal{C}^3_{rnn}\right)\right]. \label{longituidinal_A1_C}
\end{align}

\noindent Substituting the above expression in Eq.~\eqref{1_3law} and integrating subsequently, we get
\begin{align}
    \mathcal{C}^1_{rrr} + \xi\left(\mathcal{C}^2_{rrr} + 2\mathcal{C}^3_{rrr} \right) - \frac{\xi}{r^4}\int 2r^3 \frac{\partial }{\partial r}\left[r\left(\mathcal{C}^3_{rrr} + 2\mathcal{C}^3_{rnn}\right)\right]dr &= - \frac{2}{15}\varepsilon r.\label{2_by_15_law} 
\end{align}
Finally, based on the Eqs.~\eqref{C12_rrr_C12_nnr}--\eqref{B2_rnn}, Eq.~\eqref{2_by_15_law} can be expressed in terms of the structure functions leading to the following form -
\begin{equation}
\mathcal{B}^1_{rrr} + 3 \xi \mathcal{B}^2_{rrr} - \frac{3\xi}{r^4}\int r^3 \frac{\partial }{\partial r}\left[r\left(\mathcal{B}^2_{rrr} + 2\mathcal{B}^2_{rnn}\right)\right]dr = - \frac{4}{5}\varepsilon r. \label{4_5law}
\end{equation}

\noindent Eqs.~\eqref{2_by_15_law} and \eqref{4_5law} represent the CHNS analog of the $2/15$ and $4/5$ exact laws of hydrodynamic turbulence, respectively. In addition to the purely kinetic contribution, these exact laws also include contributions from the interfacial gradients. 
Unlike the original $2/15$ and $4/5$ exact laws in turbulence, which are written purely in terms of longitudinal correlators or longitudinal structure functions, here each of the exact laws includes a non-longitudinal contribution.  

In the following, we numerically investigate the validity of our derived scaling laws. We will present a systematic comparison of homogeneous and isotropic forms, as well as the correlator and structure functions forms of various exact laws.


\section{Numerical investigation of isotropic exact laws}\label{numerical_investigation_of_exact_laws}
\subsection{Simulation details}
\vspace{-0.5cm}
 We perform three-dimensional DNSs with grid resolutions $512^3$ (Run1) and $1024^3$ (Run2), respectively (See Tab.~\ref{Table1} for numerical parameter). The Eqs.~\eqref{Eu}-\eqref{EQ} are solved using a pseudo-spectral method in a periodic box of volume $(2\pi)^3$. A $N/2$-dealiasing method (where $N$ is the number of grid points in one direction) is employed due to the presence of cubic nonlinearity in the chemical potential, which limits the maximum available wavenumber ($k_{max}$) to $N/4$. We performed parallel simulations using a Python-based message-passing interface scheme \citep{Pan_2024_Relaxation_BFT, Pan_2025_Universal_energy_cascades, Mortensen_2016_HighPerformance}. The interface width $w= 4.164 \sqrt{\kappa/|a|}$ (over which $|\phi|\leq 0.9$) is taken to be $6\Delta x$ which gives $ \kappa=(6\Delta x/4.164)^2|a|$, where $\Delta x = 2\pi/N$ is the minimum grid-spacing \citep{Jacqmin_1999_Two-PhaseNavierStokesFlowsPhase-FieldModeling,
 Pan_2025_Universal_energy_cascades}.  The mobility parameter $\mathcal{M}$ is chosen to be proportional to $\sqrt{\kappa/|a|}$ in order to maintain the effective interface force required by the surface tension $\left(\sigma = \sqrt{8 \kappa |a|^3/9b^2}\right)$ with the change of interface thickness \citep{ Magaletti_2013_sharp-inerface_limit_of_CHNS, Pan_2024_Relaxation_BFT}. 
The kinematic viscosity $\nu$ is so chosen that the Kolmogorov scale, \textit{i.e.} the smallest eddy size (${\eta} \sim k_\eta^{-1}$) is well resolved  $i.e.$ $k_{max}>k_{\eta}$. In realistic situations, the interface width is very small compared to the smallest turbulent eddies. In our simulation, to achieve reasonably high Reynolds number, we keep $k_{max}/k_{\eta}\sim 1$, which limits $w \sim \eta$ \citep{Pan_2024_Relaxation_BFT, Mukherjee_Safdari_Shardt_Kenjereš_Van_den_Akker_2019}. However, this does not affect the main objective of our study where the energy transfer is investigated across the inertial scales. 

\begin{figure}[ht!]
    \centering
\includegraphics[width=0.65\linewidth]{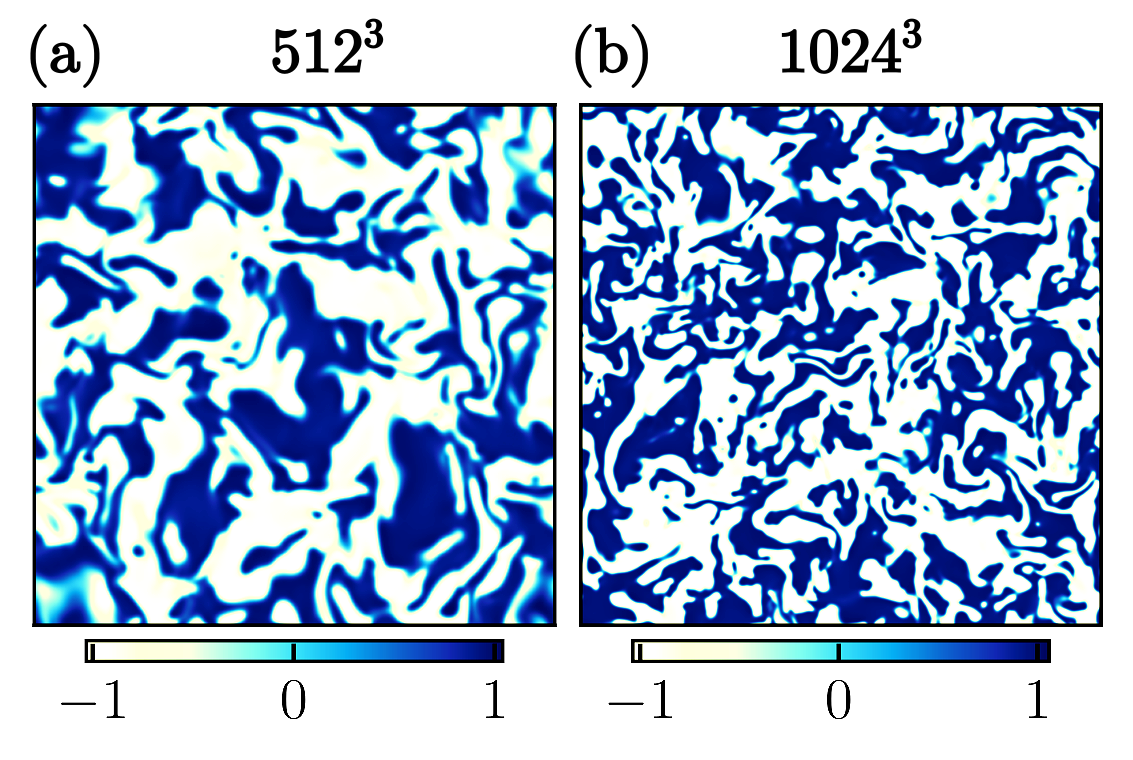}
\caption{Mid-plane snapshot of three-dimensional $\phi$ field for $512^3$ and $1024^3$ runs.}\label{snapshots_512_1024}
\end{figure}

The flow is initialized from rest ($\bu(\bm{x},0)=\bm{0}$) with a phase-mixed random distribution of the composition field ($|\phi(\bm{x},0)|\leq0.05$) and time-evolved using an unconditionally stable explicit-implicit method with an adaptive time-stepping based on the Courant–Friedrichs –Lewy (CFL) condition \citep{Eyre_1998_Unconditionally_stable_scheme,Yoon_2020_Fourier-Spectral_Method_Phase-Field_Equations}. To achieve a steady phase-locked emulsion below $T_c$, the energy is continuously pumped into the system through a large-scale Taylor green forcing $ \bm{f} \equiv f_0[sin(k_0x)cos(k_0y)cos(k_0z), - cos(k_0x)sin(k_0y)cos(k_0z),0]$,
with the forcing magnitude $f_0 = 0.5$ and $k_0 =2$. The simulations are carried out until a statistical stationary state is achieved, where the average energy injection rate is balanced by the average energy dissipation rate (see \cite{Pan_2025_Universal_energy_cascades} for details). Snapshots from such a steady state are provided for Run1 and Run2 in Fig.~\ref{snapshots_512_1024}~(a) and (b) respectively. Both figures represent phase-arrested states below $T_c$. The average domain size is in general determined by the Weber number ($We$), which measures the relative strength of the advective nonlinearity over the interfacial tension. 
A phase-arrested state with smaller domain size is therefore observed in Run2 with $We\sim 16$ in comparison with Run1 which is characterized by a smaller $We$ ($\sim 7$) (see Tab.~\ref{Table1}).

\setlength{\tabcolsep}{7pt} 
\renewcommand{\arraystretch}{1} 
\begin{table*}[ht]
\caption{{\small\label{Table1} Values of the relevant parameters (in dimensionless units) along with the three important length scales, the integral scale $L \left[= 3\pi/4 \left( \int_0^{\infty} k^{-1} E_u(k) dk/ \int_0^{\infty} E_u(k) dk\right)\right]$, the Taylor length scale $\lambda \left[= \left(5\int_0^{\infty} E_u(k) dk/\int_0^{\infty} k^{2} E_u(k) dk\right)^{1/2}\right]$, the Kolmogorov scale $\left(\eta =\left(\langle  \omega^2 \rangle/\nu^2 \right)^{1/4}\right)$, where $\bomega$ is the vorticity field and the integral-scale based Reynolds ($Re= U_L L/\nu $) and Weber ($We = U^2_L L/\sigma$) numbers where $U_{L}$ is the characteristic velocity at scale $L$ and $E_u(k)$ is the kinetic energy spectrum \cite{Andres_2019_HD_Numerical, Gomez_2000_Exact_helical_flows}. See \cite{Pan_2025_Universal_energy_cascades} for more details. }}
  \begin{tabular}{l c c c c c c c c c c c c c}
  \hline \hline 
 Runs  & $N^3$  &   $\nu$ & $\mathcal{M}$ & $\xi $ & $L$  & $\lambda$ & $\eta$  &  $L/\eta $  &  $k_{max}/k_{\eta}$ & $Re$ & $We$\\[3pt]

\hline 
Run1 & $512^3 $ & $0.0008$ & $0.010$ & $(0.0188)^2$  & $0.537$ & $0.199$ & $0.060$   & $9$ & $1.22$ & $313$ & $7$ \\

Run2 & $1024^3 $ & $0.0003$ & $0.008$ & $(0.0088)^2$  &  $0.526$ & $0.132$ & $0.029$  & $19$ & $1.19$ & $878$ & $16$\\
 \hline \hline 
 \end{tabular}
\end{table*}

\subsection{Computation of the third-order correlators}

\begin{figure}[ht!]
    \centering
\includegraphics[width=0.75\linewidth]{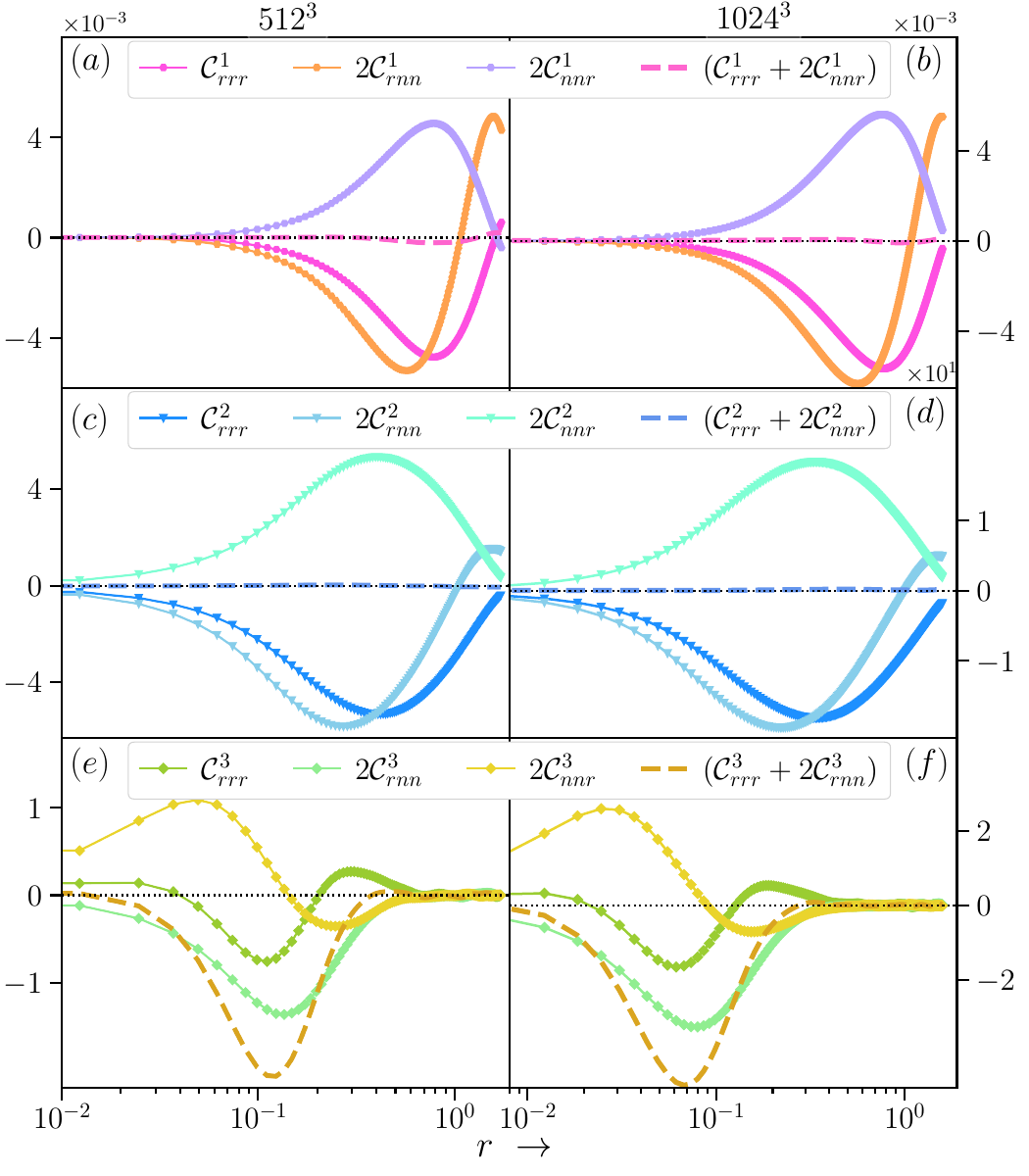}
    \caption{(a,c,e) and (b,d,f): Plots of $\mathcal{C}^{1,2,3}_{rrr}$, $\mathcal{C}^{1,2,3}_{rnn}$, $\mathcal{C}^{1,2,3}_{nnr}$ and $\mathcal{C}^{1,2,3}_{rrr}+2\mathcal{C}^{1,2,3}_{nnr}$ for Run1 and Run2, respectively. }
\label{fig:all_corrected_C_1024_512_all}
\end{figure}

We begin by computing the longitudinal velocity correlator $\mathcal{C}^1_{rrr}(r) = \langle u_r u_r u'_r \rangle$, where $u_r = \bu\cdot\widehat{\bm{r}}$ and the averaging $\langle \cdot \rangle$ consists of two parts. First, for a given $\bm{r}$, we average over all possible pairs $\bm{x}$ and $\bm{x}' = \bm{x} + \bm{r}$. This is followed by angular averaging using the method of \cite{Taylor_2003_Recovering_isotropic_statistics} over 73 directions, which yields the isotropic correlators. For computing transverse components $u_n$, we took the mean projection of $\bu$ w.r.t. two directions perpendicular to $\hat{r}$.

Starting from Eqs.~\eqref{C1_ijm}--\eqref{C3_ijm}, the theory of isotropic tensors is based on the antisymmetry property of the third-order correlators, which states that $\mathcal{C}^{1,2,3}_{ijm}(\bm{r}) = -\mathcal{C}^{1,2,3}_{ijm}(-\bm{r})$ and $\mathcal{C}^{1,2,3}_{ijm}(\bm{0}) = \bm{0}$. Numerically computing $C^{1}_{rrr}$, we found that it has a nonzero value at $r=0$, i.e., $C^{1}_{rrr}(0) = \langle u_r^3 \rangle \neq 0$. Therefore, to accommodate the analytical structure, we need to construct antisymmetrized correlators, e.g., $\mathcal{C}^{1*}_{rrr}(r)$ as
\begin{align}
\mathcal{C}^{1*}_{rrr}(r) = \frac{1}{2}\left(\langle u_r u_ru^{\prime}_r\rangle - \langle u^{\prime}_ru^{\prime}_r u_r\rangle\right) = \frac{1}{2}\left(\mathcal{C}^{1}_{rrr}(r) - \mathcal{C}^{1}_{rrr}(-r)\right).
\end{align}
Similar antisymmetrization operations are also performed on other correlators. For the sake of brevity, we drop the starred notation hereinafter, and we use the antisymmetrized correlators for all subsequent computations. In Fig.~\ref{fig:all_corrected_C_1024_512_all}, we plot $\mathcal{C}^{1,2,3}_{rrr}(r)$, $\mathcal{C}^{1,2,3}_{rnn}(r)$, $\mathcal{C}^{1,2,3}_{nnr}(r)$, and the sum $\mathcal{C}^{1,2,3}_{rrr} + 2\mathcal{C}^{1,2,3}_{rnn}$. For both runs, the correlators satisfy the relation $\mathcal{C}^{1,2}_{rrr} + 2\mathcal{C}^{1,2}_{rnn} = 0$ as predicted analytically (see the dashed lines in Figs.~\ref{fig:all_corrected_C_1024_512_all}(a), (c) and (b), (d)). The sum $\mathcal{C}^{3}_{rrr} + 2\mathcal{C}^{3}_{rnn}$, on the other hand is found to be non-zero (see Fig.~\ref{fig:all_corrected_C_1024_512_all}(e) and (f)), which is also in accordance with the analytical prediction obtained above.

\begin{figure}
    \centering
\includegraphics[width=0.85\linewidth]{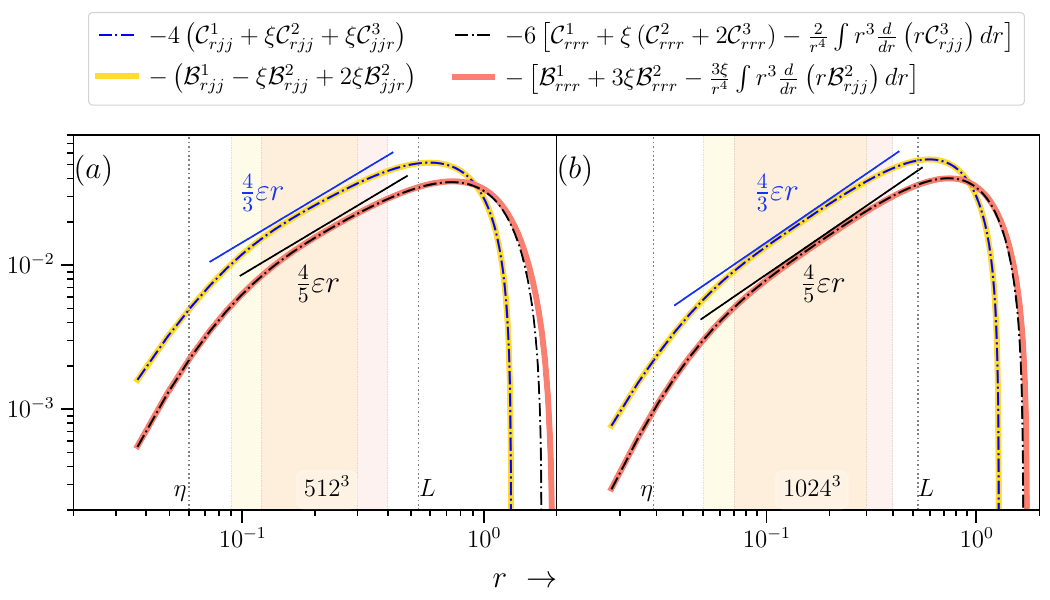}
    \caption{(a) and (b): Scaling of isotropic exact relations in the correlator form ($1/3$ and $2/15$ laws) and structure function forms ($4/3$ and $4/5$ exact laws), for Run1 and Run2, respectively. The left and right vertical lines in both the plots represent the Kolmogorov scale and the integral scales, respectively whereas the shaded region represent the regime where the linear scaling w.r.t. $r$ holds for the respective exact laws.}
    \label{fig:AC_iso_Ab_iso_4_3_4_5}
\end{figure}

\subsection{ Evidence of universal scaling}

In this section, we numerically examine various isotropic exact scaling relations obtained in terms of the correlators (the $1/3$ and $2/15$ laws) and the structure functions (the $4/3$ and $4/5$ laws). To compare the scaling of $1/3$ law \eqref{1_3law} (correlator form) with the corresponding $4/3$ law \eqref{4_3rd_law} (structure function form), we multiply the equation \eqref{1_3law} by a factor of 4 and plot them together in figure \ref{fig:AC_iso_Ab_iso_4_3_4_5}. Both Run1 and Run2 show an excellent agreement between   $-4\left(\mathcal{C}^{1}_{rjj} + \xi\mathcal{C}^{2}_{rjj} + \xi \mathcal{C}^{3}_{jjr}\right)$ (blue dashed curve) and $-\left(\mathcal{B}^{1}_{rjj} - \xi\mathcal{B}^{2}_{rjj} + 2\xi\mathcal{B}^{2}_{jjr}\right)$ (yellow curve) . Furthermore, both the quantities exhibit linear scaling with slope equal to $4\varepsilon /3$ across an intermediate range between $\eta$ and the integral scale ($L$) representing the so-called inertial range of scales (the shaded regime)  with self-similar energy cascade. Furthermore, the inertial range become wider with increase of the Reynolds number from Run1 ($Re \sim 313$) to Run2 ($Re \sim 878$) as the higher value of $Re$ entails an increased $L/\eta$ ratio.

Similarly, to compare the $2/15$ \eqref{2_by_15_law} (correlator form) and $4/5$ \eqref{4_5law} (structure function form) scaling laws, we multiply the equation \eqref{2_by_15_law} by a factor of 6 and overplot them in figure \ref{fig:AC_iso_Ab_iso_4_3_4_5}. Here too, an excellent agreement between the scaling quantities of the correlator form
\begin{equation*}
    -6\left[\mathcal{C}^{1}_{rrr} + \xi\left(\mathcal{C}^{2}_{rrr}+2\mathcal{C}^{3}_{rrr}\right) -\frac{2}{r^4}\int r^3 \frac{\partial }{\partial r}\left(r\mathcal{C}^3_{rjj}\right)dr \right]
\end{equation*}
(black dashed curve) and the structure function form 
\begin{equation*}
    -\left[\mathcal{B}^{1}_{rrr} +3 \xi \mathcal{B}^{2}_{rrr} - \frac{3\xi}{r^4}\int r^3 \frac{\partial }{\partial r}\left( r \mathcal{B}^2_{rjj}\right)dr\right]
\end{equation*} (salmon curve) is observed for both the runs.  
An excellent match between the correlator and structure function forms validates our developed vKH formulation  for isotropic BFT.  Note that slight deviations between the two forms at very large scales (beyond the integral scale) can be attributed to the lack of isotropy arising from the large-scale forcing effects.

\begin{figure}
    \centering
\includegraphics[width=0.99\linewidth]{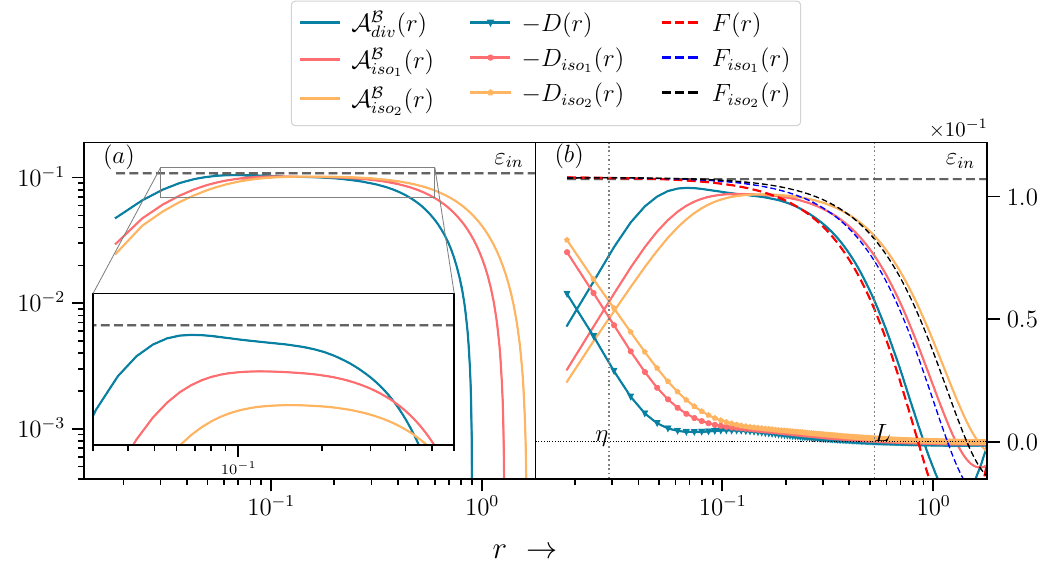}
    \caption{(a) Plots of total energy cascade rates $\mathcal{A}^{\mathcal{B}}_{div}(r)$, $\mathcal{A}^{\mathcal{B}}_{iso_1}(r)$ and $\mathcal{A}^{\mathcal{B}}_{iso_2}(r)$ in log-log scale for Run2. For better comparison, in the inset we plot a zoomed in version of the three with arbitrary shifting. (b) Plots of the corresponding two-point average energy dissipation rates $D(r)$, $D_{iso_1}(r)$ and $D_{iso_2}(r) $, two-point average energy injection rates $F(r)$, $F_{iso_1}(r)$ and $F_{iso_2}(r) $  along with the energy cascade rates in semi-log scale for Run2. The horizontal (dashed gray) lines on both the plot represent the constant energy injection rate $\varepsilon_{in}$ whereas the vertical lines on the left plot represent the Kolomogorv scale $\eta$ (left) and the integral scale $L$ (right), respectively.} 
    \label{fig:All_forms_compare_div_iso1_iso2}
\end{figure}

\subsection{Comparison of homogeneous and isotropic exact laws}

By means of DNS, we convincingly validate various exact scaling laws that we derived for isotropic binary fluid turbulence.  All these laws show a fairly linear scaling behaviour across the inertial scales where the slope of the straight line is proportional to a constant cascade rate $\varepsilon$. In theory, the constant $\varepsilon$ can be determined from any of the exact laws. In practice, however, the cascade rate $\varepsilon$ is not strictly constant but changes across the scales (including the inertial range). In such situation, it is straightforward to show that different $\varepsilon (r)$ are obtained from different exact laws. In this section, we compare $\varepsilon (r)$ obtained from the different exact laws. Earlier, we also show that the isotropic exact laws in correlator and structure function forms are in excellent agreement with each other. Therefore, for this analysis, we only use the structure function forms of the exact laws. Denoting the energy cascade rates obtained from Eqs.~\eqref{divergence_form_epsilon}, \eqref{4_3rd_law} and \eqref{4_5law} as $ \mathcal{A}^{\mathcal{B}}_{div}(r)$, $\mathcal{A}^{\mathcal{B}}_{iso_{1}}(r)$ and $\mathcal{A}^{\mathcal{B}}_{iso_{2}}(r)$, respectively, we find the following hierarchical relation among them  
\begin{align}
\mathcal{A}^{\mathcal{B}}_{iso_1}(r) &=  \frac{3}{r^3} \int \mathcal{A}^{\mathcal{B}}_{div}(r) r^2 dr, \,\text{and}\,\,
\mathcal{A}^{\mathcal{B}}_{iso_2}(r) =  \frac{5}{r^5} \int \mathcal{A}^{\mathcal{B}}_{iso_1}(r) r^4 dr.\label{A_C,B_iso1_to_C,B_iso2}
\end{align}
 In Fig.~\ref{fig:All_forms_compare_div_iso1_iso2} (a), we plot $\mathcal{A}^{\mathcal{B}}_{div}(r)$, $\mathcal{A}^{\mathcal{B}}_{iso_1}(r)$ and $\mathcal{A}^{\mathcal{B}}_{iso_2}(r)$ for Run2. For all three cases, a universal cascade regime is found where the constant cascade rate touches the energy injection rate (the dashed gray line). However, upon further zooming, we find that the corresponding inertial ranges are not perfectly flat but has an increasing flatness as one goes from the divergence form, to the $4/3$ law, and finally to the $4/5$ law (see the inset of figure \ref{fig:All_forms_compare_div_iso1_iso2} (a)). As noted in Eq.~\eqref{A_C,B_iso1_to_C,B_iso2}, the divergence form provides the cascade rate without any integration, whereas $\mathcal{A}_{iso_1}$ is obtained by integrating $\mathcal{A}_{div}$, leading to a comparatively smoother curve of $\mathcal{A}^{\mathcal{B}}_{iso_1}$ than $\mathcal{A}^{\mathcal{B}}_{div}$. Following similar arguments, one can further conclude that $\mathcal{A}^{\mathcal{B}}_{iso_2}(r)$ is smoother (or flatter) compared to $\mathcal{A}^{\mathcal{B}}_{iso_1}(r)$. We also found that the associated inertial ranges are shifted from each other. In particular, there is an infrared shift of the inertial range as one goes from homogeneous to the isotropic case (see the inset plots of Fig.~\ref{fig:All_forms_compare_div_iso1_iso2} (a)). This is also consistent with the shift of the scaling regimes observed in Fig.~\ref{fig:AC_iso_Ab_iso_4_3_4_5}, where the inertial range of the $4/5$ law (light salmon region) is shifted to larger scales compared with that of the $4/3$ law (light yellow region). 
To explain these, note that the two-point average energy injection and dissipation rates are also related by the integration hierarchy (see Eq.~\eqref{vKH_BFT}):
\begin{align}
 F_{iso_1}(r) =  \frac{3}{r^3} \int F(r) r^2 dr \,\,\,\,&\text{and}\,\,\,\,F_{iso_2}(r) =  \frac{5}{r^5} \int F_{iso_1}(r) r^4 dr,\label{FD_B_div_B_iso1}\\
D_{iso_1}(r) =  \frac{3}{r^3} \int D (r) r^2 dr\,\,\,&\text{and}\,\,\,\, D_{iso_2}(r) =  \frac{5}{r^5} \int D_{iso_1}(r) r^4 dr\label{DD_B_iso1_to_B_iso2}.
\end{align}
Due to the successive integrations involved in the forcing and dissipative terms, the small-scale energy fluctuations and dissipative effects are smoothed out and therefore they start to appear at comparatively larger scales. This can be seen from the successive shift of $F(r)$, $F_{{iso}_1}$, and $F_{{iso}_2}$ (dashed curves), and $D(r)$, $D_{{iso}_1}$, and $D_{{iso}_2}$ (symbolic curves), toward larger scales in Fig.~\ref{fig:All_forms_compare_div_iso1_iso2}(b). Consequently, the flux terms $\mathcal{A}^{\mathcal{B}}_{div}$, $\mathcal{A}^{\mathcal{B}}_{{iso}_1}$, and $\mathcal{A}^{\mathcal{B}}_{{iso}_2}$ also exhibit an infrared shift in order to maintain the stationarity. Overall, we argue that isotropization by constitutive integrations acts as a low-pass filter that shifts the inertial range to larger scales and results in flatter cascade regimes, providing $\varepsilon(r)$ that more closely represent Kolmogorov's hypothesis of a scale-independent inertial-range cascade.

\section{Summary and conclusion}\label{summary_and_conclusion}
Kolmogorov's classic $4/5$ law is the most fundamental scaling relation of isotropic turbulence 
which yields the linear scaling of the third-order longitudinal velocity structure function. However, whether such a scaling law also holds in the presence of interfacial dynamics in phase-arrested CHNS turbulence has remained an open question. In this article, we address this fundamental problem in a systematic manner. Using the vKH formulation of isotropic tensors, we first derive the $1/3$ and $4/3$ exact scaling relations for total energy transfer in CHNS turbulence, expressed in terms of third-order correlators and structure functions, respectively. Subsequently, further algebraic manipulations and incorporating  the curl-free nature of $\mathbf{Q}$, we finally obtain the isotropic $2/15$ law (in correlator form) and the corresponding $4/5$ law (in structure function form), both containing contributions from kinetic and interfacial dynamics.

\noindent Using DNS with resolutions up to $1024^3$ grid points, we show that the correlator and structure-function formulations are equivalent, and all the isotropic exact laws exhibit Kolmogorov-type linear scaling across the inertial range of scales. These results therefore justify the proposed vKH framework for isotropic CHNS turbulence in the presence of interfacial dynamics and provide the flexibility to compute the constant cascade rate $\varepsilon$ from any of the exact laws. In practice, however, we find that $\varepsilon$ is not strictly constant across the inertial range but instead exhibits a weak dependence on the separation scale $r$. In particular, $\varepsilon(r)$ obtained from the various exact laws is connected through a hierarchy of integrations, progressing successively from the homogeneous to the isotropic forms. The process of isotropization through integration increases the smoothness, or flatness, of $\varepsilon(r)$ and consequently, the cascade rate obtained from the $4/5$ law more closely resembles Kolmogorov’s phenomenology of universal scale-independent cascade. Furthermore, this isotropization acts as a low-pass filter that smooths out small-scale fluctuations, resulting in an infrared shift of the associated inertial range. This study provides a fundamental step forward to probe Kolmogorov's refined similarity hypothesis where the scaling laws are defined with respect to the locally averaged energy $\varepsilon(r)$ rather than the global constant $\varepsilon$ \citep{Kolmogorov_1962, Stolovitzky_1992_refined_similarity, Yao_2024_PRL_refined_hypothesis}. \\

Apart from enabling the computation of the energy cascade rate or the turbulent heating rate, the importance of the $4/5$ law for isotropic turbulence is prominently acknowledged to probe the nature of intermittency in turbulent flows \citep{frisch1995turbulence}. Our derived $4/5$ law may similarly provide the correct scaling quantity for this purpose for CHNS turbulence. However, unlike the purely longitudinal velocity structure functions in the NS case, the $4/5$ exact law for CHNS turbulence consists of a sum of a purely longitudinal contribution, $\mathcal{S}_{\|}$, and an additional mixed contribution involving both longitudinal and transverse components, $\mathcal{S}^{a}$. A separate study is thus necessary to understand the mutual importance of various terms in the exact scaling law for varying Weber numbers and therefore to investigate the role of interfacial coupling in CHNS turbulence and the corresponding intermittency \citep{Ray_PRE_2011_scaling_and_multiscaling_in_turbulent_BFT}. Furthermore, our work provides a strong theoretical framework for the study of isotropic turbulence in systems that do not necessarily involve divergence-free variables, such as multiphase flows (e.g., bubbly flows and droplet-laden turbulence), as well as other complex systems, including polymeric and activity-driven fluids, ferrofluids, and related systems \citep{Esposito_2023_intermittency_in_turbulent_emulsion, Mangani_2024_drop_laden_turbulence,
Ma_2025_Kolmogorov_scaling_Bubble-Induced,   Bhattacharjee_2021_driven_active_matter, Chiarini_2025_polymeric_turbulence, Mouraya_2019_ferrohydrodynamic_turbulence, Mouraya_2024_Staionary_nonstationary_cascades}. Finally, the role of Hinze scale and subsequent possibility of a nonlocal energy cascade in CHNS turbulence is also a future direction to explore \cite{halder2025localnonlocalenergytransfers}.

\section{Acknowledgment}
The simulations are performed using the support and resources provided by PARAM Sanganak under the National Supercomputing Mission (NSM), Government of India, at the Indian Institute of Technology, Kanpur. For data analysis, we again acknowledge NSM for providing computing resources
resources of ‘PARAM UTKARSH’ at CDAC Knowledge Park, Bangalore, which is implemented
by C-DAC and supported by the Ministry of Electronics and Information Technology (MeitY)
and Department of Science and Technology (DST), Government of India.

\end{document}